# On the Search for High-Rate Quasi-Orthogonal Space-Time Block Code

Chau Yuen, Yong Liang Guan, Tjeng Thiang Tjung

**Abstract –** A Quasi-Orthogonal Space-Time Block Code (QO-STBC) is attractive because it achieves higher code rate than Orthogonal STBC and lower decoding complexity than non-orthogonal STBC. In this paper, we first derive the algebraic structure of QO-STBC, then we apply it in a novel graph-based search algorithm to find high-rate QO-STBCs with code rates greater than 1. From the four-antenna codes found using this approach, it is found that the maximum code rate is limited to 5/4 with symbolwise diversity level of four, and 4 with symbolwise diversity level of two. The maximum likelihood decoding of these high-rate QO-STBCs can be performed on two separate sub-groups of symbols. The rate-5/4 codes are the first known QO-STBCs with code rate greater 1 that has full symbolwise diversity level.

## 1. INTRODUCTION

Orthogonal Space-Time Block Code (O-STBC) can provide full transmit diversity with simple linear maximum-likelihood (ML) decoding complexity. This, however, limits its maximum achievable code rate to be less than one when the number of transmit antennas exceeds two [1]. Specifically, for 4 transmit antennas, the maximum achievable code rate of O-STBC has been shown to be 3/4. From the information theoretic viewpoint [2], this implies that O-STBC suffers a loss of capacity. As a result, Quasi-Orthogonal Space-Time Block Code (QO-STBC), which can achieve a higher code rate than O-STBC at the expense of a slightly higher decoding complexity, has been investigated [3-7]. The ML decoding of QO-STBC can be achieved by jointly detecting a sub-group of the transmitted symbols, rather than all the transmitted symbols, hence QO-STBC leads to a lower decoding complexity than general non-orthogonal STBC.

In [6,7], it is shown that QO-STBC for four transmit antennas that require the joint detection of only two real symbols for ML decoding has a maximum code rate of 1. This, when compared against the maximum code rate of 3/4 and linear decoding complexity for an O-STBC for the same number of transmit antennas, suggests that the maximum code rate of QO-STBC may exceed 1 if a decoding complexity higher than the joint detection of two real symbols is permitted. In this paper, we set out to search for such "high-rate QO-STBC" with code rate greater than one. To our knowledge, this is the first such attempt ever conducted on QO-STBC.

The organization of this paper is as follows: Section 2 gives the signal model and derives the unifying algebraic structure of a QO-STBC. Section 3 discusses the modeling of the code search problem, and introduces a novel graph-based algorithm to perform the code search, based on graph theory. Finally Section 4 concludes the paper.

## 2. QUASI-ORTHOGONAL SPACE-TIME BLOCK CODE

*2.1 STBC Model*

Suppose that there are $N_t$ tansmit antennas, $N_r$ receive antennas, and an interval of $T$ symbol periods during which the propagation channel is constant and known to the receiver. The information data sequence is segmented into blocks of $K$ complex symbols $\{x_1, x_2, ..., x_K\}$, each for transmission using a STBC. The transmitted STBC codeword can be written as a $T \times N_t$ matrix $\mathbf{G}$ that governs the transmission over $N_t$ antennas and $T$ symbol periods. The code rate of the STBC is defined as $K/T$. Following the STBC model in [8] with the complex symbol $x_k$ expressed as $x_k = s_k + js_{K+k}$, $1 \leq k \leq K$, the STBC codeword $\mathbf{G}$ can be expressed as:

$$\mathbf{G} = \sum_{i=1}^{2K}(s_i \mathbf{A}_i) \qquad (1)$$

where the matrices $\mathbf{A}_i$ are called the "dispersion matrices", have size $T \times N_t$, and are normalised by the power distribution constraint $\text{tr}(\mathbf{A}_i^H \mathbf{A}_i) = TN_t / K$ [8]. The received signal model for a system with multiple receive antennas can been shown to be:

$$\tilde{\mathbf{r}} = \sqrt{\rho/N_t}\,\mathbf{H}\tilde{\mathbf{s}} + \tilde{\boldsymbol{\eta}} \qquad (2)$$

where the normalization $\sqrt{\rho/N_t}$ is to ensure that the SNR ($\rho$) at the receiver is independent of the number of transmit antennas, and

$$\tilde{\mathbf{r}} = \begin{bmatrix} \mathbf{r}_1^R \\ \mathbf{r}_1^I \\ \vdots \\ \mathbf{r}_{N_r}^R \\ \mathbf{r}_{N_r}^I \end{bmatrix}, \tilde{\mathbf{s}} = \begin{bmatrix} s_1 \\ \vdots \\ s_K \\ s_{K+1} \\ \vdots \\ s_{2K} \end{bmatrix}, \tilde{\boldsymbol{\eta}} = \begin{bmatrix} \boldsymbol{\eta}_1^R \\ \boldsymbol{\eta}_1^I \\ \vdots \\ \boldsymbol{\eta}_{N_r}^R \\ \boldsymbol{\eta}_{N_r}^I \end{bmatrix},$$

$$\mathbf{H} = \begin{bmatrix} \mathcal{A}_1\underline{\mathbf{h}}_1 & \mathcal{A}_2\underline{\mathbf{h}}_1 & \cdots & \mathcal{A}_K\underline{\mathbf{h}}_1 & \cdots & \mathcal{A}_{2K}\underline{\mathbf{h}}_1 \\ \vdots & \vdots & \ddots & \vdots & \ddots & \vdots \\ \mathcal{A}_1\underline{\mathbf{h}}_{N_r} & \mathcal{A}_2\underline{\mathbf{h}}_{N_r} & \cdots & \mathcal{A}_K\underline{\mathbf{h}}_{N_r} & \cdots & \mathcal{A}_{2K}\underline{\mathbf{h}}_{N_r} \end{bmatrix}, \mathcal{A}_i = \begin{bmatrix} \mathbf{A}_i^R & -\mathbf{A}_i^I \\ \mathbf{A}_i^I & \mathbf{A}_i^R \end{bmatrix}, \underline{\mathbf{h}}_i = \begin{bmatrix} \mathbf{h}_i^R \\ \mathbf{h}_i^I \end{bmatrix}.$$

In (2), $\mathbf{r}_i$ and $\boldsymbol{\eta}_i$, where $1 \leq i \leq N_r$, are $T \times 1$ column vectors which contain the received signal and zero-mean unit-variance AWGN noise for the $i^{th}$ receive antenna over $T$ symbol periods; $\mathbf{H}$ is the equivalent channel matrix, and $\mathbf{h}_i$ is a $N_t \times 1$ column vector that contains the Rayleigh flat fading coefficients from the $N_t$ transmit antennas to $i^{th}$ receive antenna. The superscripts $(\ )^R$ and $(\ )^I$ denote the real and imaginary parts respectively of a complex element, vector or matrix.

## 2.2 Algebraic Structure of QO-STBC

The concept of quasi-orthogonality in QO-STBC refers to the division of the $K$ transmitted symbols in a QO-STBC codeword into $G$ different groups such that symbols in one group are orthogonal to all symbols in the other groups, while strict orthogonality among the symbols within a group is not required. Orthogonality in this case means that, by linear matched filtering at the receiver, the received symbols of the QO-STBC can be de-coupled into $G$ independent groups, and the ML decoding of different groups can be performed separately and in parallel by jointly detecting only $K/G$ complex symbols within a group [9], instead of jointly detecting all $K$ complex symbols within a codeword (which is clearly a more complex operation).

*Definition 1*: A quasi-orthogonal design is such that its corresponding $\mathbf{H}^T\mathbf{H}$ (with $\mathbf{H}$ as defined in (2)) can be rearranged into a block-diagonal matrix with non-zero sub-matrices of size $(2K/G)$

× (2*K*/*G*) by a permutation, i.e. $\mathbf{P}^{\mathrm{T}}\mathbf{H}^{\mathrm{T}}\mathbf{H}\mathbf{P}$ is block diagonal, where $\mathbf{P}^{\mathrm{T}}\mathbf{P} = \mathbf{I}$ and $\mathbf{P}$ has only unit entries.

Without loss of generality, throughout this paper, we assume that $\mathbf{P} = \mathbf{I}$, where $\mathbf{I}$ is an identity matrix of appropriate dimension. It should be noted that, in contrast to *Definition 1*, an *orthogonal* design requires the $\mathbf{H}^{\mathrm{T}}\mathbf{H}$ to be a scaled identity matrix [1], instead of a block-diagonal matrix for the *quasi-orthogonal* design.

In order to separate the received signals into *G* orthogonal groups, a matched filter $\mathbf{H}^{\mathrm{T}}$ is multiplied to the received signal $\tilde{\mathbf{r}}$ in (2). Let us consider a snapshot of $\mathbf{H}^{\mathrm{T}}\mathbf{H}$ as follows:

$$\mathbf{H}^{\mathrm{T}}\mathbf{H} = \begin{bmatrix} \vdots & \cdots & \vdots \\ \underline{\mathbf{h}}_1^{\mathrm{T}}\mathcal{A}_p^{\mathrm{T}} & \cdots & \underline{\mathbf{h}}_{N_r}^{\mathrm{T}}\mathcal{A}_p^{\mathrm{T}} \\ \underline{\mathbf{h}}_1^{\mathrm{T}}\mathcal{A}_u^{\mathrm{T}} & \cdots & \underline{\mathbf{h}}_{N_r}^{\mathrm{T}}\mathcal{A}_u^{\mathrm{T}} \\ \underline{\mathbf{h}}_1^{\mathrm{T}}\mathcal{A}_q^{\mathrm{T}} & \cdots & \underline{\mathbf{h}}_{N_r}^{\mathrm{T}}\mathcal{A}_q^{\mathrm{T}} \\ \underline{\mathbf{h}}_1^{\mathrm{T}}\mathcal{A}_v^{\mathrm{T}} & \cdots & \underline{\mathbf{h}}_{N_r}^{\mathrm{T}}\mathcal{A}_v^{\mathrm{T}} \\ \vdots & \cdots & \vdots \end{bmatrix} \begin{bmatrix} \cdots & \mathcal{A}_p\underline{\mathbf{h}}_1 & \mathcal{A}_u\underline{\mathbf{h}}_1 & \mathcal{A}_q\underline{\mathbf{h}}_1 & \mathcal{A}_v\underline{\mathbf{h}}_1 & \cdots \\ \vdots & \vdots & \vdots & \vdots & \vdots & \vdots \\ \cdots & \mathcal{A}_p\underline{\mathbf{h}}_{N_r} & \mathcal{A}_u\underline{\mathbf{h}}_{N_r} & \mathcal{A}_q\underline{\mathbf{h}}_{N_r} & \mathcal{A}_v\underline{\mathbf{h}}_{N_r} & \cdots \end{bmatrix}$$

$$= \begin{bmatrix} \cdots & \cdots & \cdots & \cdots & \cdots & \cdots \\ \vdots & \sum_{i=1}^{N_R}\underline{\mathbf{h}}_i^{\mathrm{T}}(\mathcal{A}_p^{\mathrm{T}}\mathcal{A}_p)\underline{\mathbf{h}}_i & \sum_{i=1}^{N_R}\underline{\mathbf{h}}_i^{\mathrm{T}}(\mathcal{A}_p^{\mathrm{T}}\mathcal{A}_u)\underline{\mathbf{h}}_i & \boxed{\sum_{i=1}^{N_R}\underline{\mathbf{h}}_i^{\mathrm{T}}(\mathcal{A}_p^{\mathrm{T}}\mathcal{A}_q)\underline{\mathbf{h}}_i} & \sum_{i=1}^{N_R}\underline{\mathbf{h}}_i^{\mathrm{T}}(\mathcal{A}_p^{\mathrm{T}}\mathcal{A}_v)\underline{\mathbf{h}}_i & \vdots \\ \vdots & \sum_{i=1}^{N_R}\underline{\mathbf{h}}_i^{\mathrm{T}}(\mathcal{A}_u^{\mathrm{T}}\mathcal{A}_p)\underline{\mathbf{h}}_i & \sum_{i=1}^{N_R}\underline{\mathbf{h}}_i^{\mathrm{T}}(\mathcal{A}_u^{\mathrm{T}}\mathcal{A}_u)\underline{\mathbf{h}}_i & \sum_{i=1}^{N_R}\underline{\mathbf{h}}_i^{\mathrm{T}}(\mathcal{A}_u^{\mathrm{T}}\mathcal{A}_q)\underline{\mathbf{h}}_i & \sum_{i=1}^{N_R}\underline{\mathbf{h}}_i^{\mathrm{T}}(\mathcal{A}_u^{\mathrm{T}}\mathcal{A}_v)\underline{\mathbf{h}}_i & \vdots \\ \vdots & \boxed{\sum_{i=1}^{N_R}\underline{\mathbf{h}}_i^{\mathrm{T}}(\mathcal{A}_q^{\mathrm{T}}\mathcal{A}_p)\underline{\mathbf{h}}_i} & \sum_{i=1}^{N_R}\underline{\mathbf{h}}_i^{\mathrm{T}}(\mathcal{A}_q^{\mathrm{T}}\mathcal{A}_u)\underline{\mathbf{h}}_i & \sum_{i=1}^{N_R}\underline{\mathbf{h}}_i^{\mathrm{T}}(\mathcal{A}_q^{\mathrm{T}}\mathcal{A}_q)\underline{\mathbf{h}}_i & \sum_{i=1}^{N_R}\underline{\mathbf{h}}_i^{\mathrm{T}}(\mathcal{A}_q^{\mathrm{T}}\mathcal{A}_v)\underline{\mathbf{h}}_i & \vdots \\ \vdots & \boxed{\sum_{i=1}^{N_R}\underline{\mathbf{h}}_i^{\mathrm{T}}(\mathcal{A}_v^{\mathrm{T}}\mathcal{A}_p)\underline{\mathbf{h}}_i} & \sum_{i=1}^{N_R}\underline{\mathbf{h}}_i^{\mathrm{T}}(\mathcal{A}_v^{\mathrm{T}}\mathcal{A}_u)\underline{\mathbf{h}}_i & \sum_{i=1}^{N_R}\underline{\mathbf{h}}_i^{\mathrm{T}}(\mathcal{A}_v^{\mathrm{T}}\mathcal{A}_q)\underline{\mathbf{h}}_i & \sum_{i=1}^{N_R}\underline{\mathbf{h}}_i^{\mathrm{T}}(\mathcal{A}_v^{\mathrm{T}}\mathcal{A}_v)\underline{\mathbf{h}}_i & \vdots \\ \cdots & \cdots & \cdots & \cdots & \cdots & \cdots \end{bmatrix} \quad (3)$$

where $1 \leq p, q, u, v \leq 2K$.

Assume that the symbols $s_p$ and $s_u$ are in the same group (hence they are not orthogonal), while the symbols $s_q$ and $s_v$ are in another group (hence they are orthogonal to $s_p$ and $s_u$). We write {*p*, *u*} ⊂ $\mathcal{G}(p)$ and {*q*, *v*} ⊄ $\mathcal{G}(p)$, where $\mathcal{G}(p)$ represents a set of symbol indices that are in the same group as symbol with index $p$, including $p$; similarly, $\{q, v\} \subset \mathcal{G}(q)$ and $\{p, u\} \not\subset \mathcal{G}(q)$. In order to achieve orthogonality among the symbols of different groups, e.g. between symbols $s_u$ and $s_v$, the summation terms included in the boxes in (3) are required to be zero, hence $\mathcal{A}_u^T \mathcal{A}_v$ and $\mathcal{A}_v^T \mathcal{A}_u$ (likewise for $\mathcal{A}_p^T \mathcal{A}_v$, $\mathcal{A}_u^T \mathcal{A}_q$, $\mathcal{A}_p^T \mathcal{A}_q$, etc.) have to be skew-symmetric, i.e. $(\mathcal{A}_u^T \mathcal{A}_v)^T = -\mathcal{A}_u^T \mathcal{A}_v$, $(\mathcal{A}_v^T \mathcal{A}_u)^T = -\mathcal{A}_v^T \mathcal{A}_u$ and so on.

*Theorem 1*: For symbols with index $u$ and $v$ to be orthogonal to each other, i.e. $\mathcal{A}_u^T \mathcal{A}_v$ and $\mathcal{A}_v^T \mathcal{A}_u$ to be skew-symmetry, the following *Quasi-Orthogonality Constraint* (QOC) has to be fulfilled:

$$\mathbf{A}_u^H \mathbf{A}_v = -\mathbf{A}_v^H \mathbf{A}_u \quad 1 \leq u, v \leq 2K \quad \text{and} \quad v \notin \mathcal{G}(u) \tag{4}$$

Proof of *Theorem 1*:

$$\begin{aligned} &\mathbf{A}_u^H \mathbf{A}_v = -\mathbf{A}_v^H \mathbf{A}_u \\ &\Rightarrow \left(\mathbf{A}_u^R + j\mathbf{A}_u^I\right)^H \left(\mathbf{A}_v^R + j\mathbf{A}_v^I\right) = -\left(\mathbf{A}_v^R + j\mathbf{A}_v^I\right)^H \left(\mathbf{A}_u^R + j\mathbf{A}_u^I\right) \\ &\Rightarrow \begin{cases} \text{real part: } \left(\mathbf{A}_u^R\right)^T \mathbf{A}_v^R + \left(\mathbf{A}_u^I\right)^T \mathbf{A}_v^I = -\left(\mathbf{A}_v^R\right)^T \mathbf{A}_u^R - \left(\mathbf{A}_v^I\right)^T \mathbf{A}_u^I \\ \text{imag part: } \left(\mathbf{A}_u^R\right)^T \mathbf{A}_v^I - \left(\mathbf{A}_u^I\right)^T \mathbf{A}_v^R = -\left(\mathbf{A}_v^R\right)^T \mathbf{A}_u^I + \left(\mathbf{A}_v^I\right)^T \mathbf{A}_u^R \end{cases} \end{aligned} \tag{5}$$

Defining

$$\Rightarrow \begin{cases} \mathbf{M} \triangleq \left(\mathbf{A}_u^R\right)^T \mathbf{A}_v^R + \left(\mathbf{A}_u^I\right)^T \mathbf{A}_v^I \\ \mathbf{N} \triangleq \left(\mathbf{A}_u^R\right)^T \mathbf{A}_v^I - \left(\mathbf{A}_u^I\right)^T \mathbf{A}_v^R \end{cases}$$

then (5) shows that **M** is skew-symmetry (i.e. $\mathbf{M}^T = -\mathbf{M}$) and **N** is symmetric (i.e. $\mathbf{N}^T = \mathbf{N}$). Since $\mathcal{A}_u^T \mathcal{A}_v = \begin{bmatrix} \mathbf{M} & -\mathbf{N} \\ \mathbf{N} & \mathbf{M} \end{bmatrix}$, it is easy to verify that $\mathcal{A}_u^T \mathcal{A}_v$ is skew-symmetry. Similar conclusion can be shown for $\mathcal{A}_v^T \mathcal{A}_u$. Therefore, by restricting the dispersion matrices **A** of symbols belong to different groups to satisfy the QOC in (4), orthogonality among the symbols of different groups can be achieved. Hence *Theorem 1* is proved. ∎

Note that when there is only one real symbol in a group, the QOC in *Theorem 1* becomes the orthogonality constraint for O-STBC provided in [1,14], and the condition $v \notin \mathcal{G}(u)$ in (4) is no longer needed as there will be only one real symbol in a group for O-STBC.

It can be shown that all the dispersion matrices of the QO-STBCs in [3-7] conform to the QOC in (4), hence (4) formulates the algebraic structure of a generic QO-STBC. We will make use of this algebraic structure to search for dispersion matrices that form a QO-STBC with the desired code rate in the next section.

## 3. SEARCH OF HIGH RATE QO-STBC

### 3.1 Code Parameters

Before performing the code search, we first define the parameters used for generating the dispersion matrices of a QO-STBC:

- *Code length*: The code length, *T*, of the QO-STBC is assumed be equal to the number of

transmit antennas, i.e. $T = N_t$ in this paper. So the codeword is a square matrix.

- *Matrix Entries*: The entries of the dispersion matrix are set to be $\{0, \pm 1, \pm j\}$, as commonly adopted in the literature [1,3-7].

- *Matrix Rank*: The rank of a dispersion matrix is related to the symbolwise diversity of the STBC, e.g. a code with dispersion matrices of rank 2 can never provide transmit diversity greater than 2. A dispersion matrix with rank equal to $N_t$ is said to achieve the maximal symbolwise diversity [10].

- *Matrix Weight*: This parameter refers to the number of non-zero entries in every row of a dispersion matrix.

- *Group*: This parameter is as defined in Section 2.2 and *Definition* 1. For QO-STBC, $G$ is required to be more than 1 ($G = 1$ gives a fully non-orthogonal STBC).

*3.2 Code Search Methodology*

Our proposed code search methodology works as follows: we first construct a set of "seed matrices" as potential QO-STBC dispersion matrices, and then perform a search among these matrices to look for valid QO-STBC dispersion matrices. In order to increase the chance of finding QO-STBC with rate greater than 1, it is desired to have as many seed matrices as possible. Hence in this paper, we consider seed matrices with complex number entries, as well as matrices with both weights 1 and 2.

The code search procedure can be broadly formulated into three steps:

(a) Generate a series of $N$ matrices (the so called "seed matrices") with desired parameters, e.g. $T = N_t = 4$, rank = 4, weight = 2.

(b) From the matrices obtained in Step (a), identify and retain those that can be grouped into $G$ groups according to the QOC in (4).

(c) From the matrices obtained in Step (b), identify those that can provide a code rate greater than one, i.e. matrices that give an equivalent channel matrix of rank $> 2T$ (relationship between the rank of the equivalent channel matrix and the code rate of a QO-STBC will be explained in the following part).

The above steps are elaborated below.

Step (a)

Consider the case of $T = N_t = 4$, rank 4 and weight 2. To generate matrices with these parameters, we start with the sixteen 2×2 complex Hadamard matrices with entries $\{0, \pm 1, \pm j\}$ shown in Figure 1, and use them as the **P** and **Q** sub-matrices in the sixteen 4×4 matrices shown in Figure 2. By doing so, $N = 16^3 = 4096$ matrices of size 4×4, rank 4 and weight 2 can be generated.

Step (b)

Group these N = 4096 matrices into $G$ groups based on the QOC in (4). However this is an NP-complete problem because each of the $N$ matrices can be either in one of the $G$ groups, or not in any group at all, so there are $(G+1)^N$ possible combinations. Hence an efficient algorithm is required to expedite this search/grouping process.

A novel method based on graph theory is proposed in this paper to accomplish this task. It will be described in Section 3.3.

Step (c)

Assume that $M$ dispersion matrices (grouped into $G$ groups) are found in Step (b). The objective of Step (c) is to check the rank, $R$, of the equivalent channel matrix **H** (2) formed by these $M$ dispersion matrices. $R$ represents the number of real symbols that can be supported by the equivalent channel **H**. If $R < M$, it implies that $M - R$ dispersion matrices are linearly dependent on the $R$ independent dispersion matrices. So only $R$ out of the $M$ dispersion matrices can be used to form a QO-STBC with a resultant code rate of $R/(2T)$. On the other hand, $R = M$ is the maximum achievable value for $R$. In order to achieve a code rate greater than one, it is required that $R > 2T$.

*3.3 Graph Modelling and Modified Depth First Search for Implementing Step (b)*

Now we present an efficient graph-based technique to identify the matrices from those found in Step (a) that can be grouped into $G$ groups according to the QOC in (4). The quasi-orthogonal relationship between the set of $N = 4096$ matrices can be visualized in Figure 3, which indicates the matrix index $u$ on the x-axis and the matrix index $v$ on the y-axis for all $1 \leq u, v \leq N$, and marks the $(u, v)$ point as a dark pixel if the corresponding $(\mathbf{A}_u, \mathbf{A}_v)$ matrices satisfy the QOC. A close examination of Figure 3 shows that every matrix satisfies the QOC with another 56 matrices. So if we view Figure 3 as a matrix, it is a sparse matrix in which only $56/4096 = 1.37\%$ of the matrix entries are non-zero (this is much less than the general definition of sparse matrix which require non-zero entries to be less than 10%) [11].

One of the efficient representations of sparse matrix is the graph model. This suggests that the code search/grouping problem in Step (b) can be solved with the help of graph theory and graph-based algorithms. Specifically, we model the *N* matrices found in Step (a) as a series of *N* nodes in a graph. For every pair of matrices that satisfy the QOC in (4), their nodes will be connected by a unidirectional link. Hence a graph with nodes representing possible QO-STBC dispersion matrices, and connected by links denoting conformance to the QOC between the connected nodes, will be formed. A simple example of such a graph is shown in Figure 4, where the matrix $\mathbf{A}_1$ is assumed to satisfy QOC with matrices $\mathbf{A}_2$, $\mathbf{A}_4$ and $\mathbf{A}_5$; while the matrix $\mathbf{A}_3$ is assumed to satisfy QOC with matrices $\mathbf{A}_2$ and $\mathbf{A}_4$; and so on.

The graph example in Figure 4 suggests that $\mathbf{A}_1$, $\mathbf{A}_2$, $\mathbf{A}_3$ and $\mathbf{A}_4$ can form a QO-STBC with dispersion matrices $\mathbf{A}_1$ and $\mathbf{A}_3$ in a group, and dispersion matrices $\mathbf{A}_2$ and $\mathbf{A}_4$ in another group. These two groups of dispersion matrices are orthogonal to each other, because $\mathbf{A}_1$ and $\mathbf{A}_3$ establish the QOC link with $\mathbf{A}_2$ and $\mathbf{A}_4$, and vice versa. In a QO-STBC, since the dispersion matrices in *any* group must satisfy the QOC with the dispersion matrices in *all* other groups, in our proposed graph model we can always find links that connect every matrix in a group to every matrix in another group. In short, a QO-STBC forms a *fully connected* graph with nodes representing its dispersion matrices and links connecting dispersion matrices of different groups. By exploiting this property, if we randomly pick a matrix node in this graph as the starting point to perform a "spanning tree algorithm" [12], we will be able to identify the quasi-orthogonal grouping of the *N* matrices obtained from Step (a).

Depth First Search (DFS) [12] is an algorithm in graph theory that provides a systematic way to visit all the nodes in a graph from any starting node and form a spanning tree. The pseudo codes of DFS algorithm are given in Appendix A. In order to solve the dispersion matrix grouping problem in Step (b), we propose to extend the DFS algorithm to a modified DFS (MDFS) algorithm. The pseudo codes of the proposed MDFS algorithm are given in Appendix B. Essential differences between the DFS and MDFS algorithms are listed below:

- In MDFS, every node can be visited more than once. In DFS, every node can only be visited once.
- In MDFS, there is an assignment of group to the nodes visited. There is no such assignment in DFS.
- In MDFS, there is an additional requirement that every visited node must be connected with its ancestors of different groups. There is no such requirement in DFS.

Based on the graph example in Figure 4, the trees constructed by DFS and MDFS with $G = 2$ are shown in Figure 5(a) and (b) respectively. Every branch of the tree constructed by MDFS constitutes a possible solution for the dispersion matrices of a QO-STBC. For example in Figure 5(b), $\mathbf{A}_1$-$\mathbf{A}_2$-$\mathbf{A}_3$-$\mathbf{A}_4$ and $\mathbf{A}_1$-$\mathbf{A}_4$-$\mathbf{A}_6$ and $\mathbf{A}_1$-$\mathbf{A}_5$ are possible solutions, but $\mathbf{A}_1$-$\mathbf{A}_4$-$\mathbf{A}_3$-$\mathbf{A}_2$ is not as it is merely a permutation of the first branch.

On the other hand, the $\mathbf{A}_1$-$\mathbf{A}_2$-$\mathbf{A}_3$-$\mathbf{A}_4$-$\mathbf{A}_6$ branch in the DFS tree in Figure 5(a) is not a valid QO-STBC solution because although $\mathbf{A}_6$ has a QOC link with $\mathbf{A}_4$ (which is in group 2), it does not have a QOC link with its ancestor node $\mathbf{A}_2$ (which is also in group 2), hence $\mathbf{A}_6$ cannot be added as group 1 node and cannot form a QO-STBC together with $\mathbf{A}_2$ and $\mathbf{A}_4$. This explains why the basic DFS algorithm cannot be used for solving the code search problem described in Step (b).

Back to Figure 5(b), since we want a QO-STBC with as high code rate as possible, only the $\mathbf{A}_1$-$\mathbf{A}_2$-$\mathbf{A}_3$-$\mathbf{A}_4$ branch is considered as it gives a QO-STBC with the largest number of dispersion matrices. The resultant QO-STBC has a group of dispersion matrices consisting of $\mathbf{A}_1$ & $\mathbf{A}_3$, and another group of dispersion matrices consisting of $\mathbf{A}_2$ & $\mathbf{A}_4$. So a total of $M = 4$ dispersion matrices, divided into two orthogonal groups, are found. Hence the proposed MDFS algorithm can be used for solving the code search problem discussed in Step (b).

*3.4 Code Search Results*

Using the proposed MDFS algorithm with $G = 2$ on the set of $N = 4096$ matrices with rank 4 and weight 2 described earlier, we are able to find a few set of solutions each with $M = 16$ matrices grouped into two orthogonal groups. Each of these solution sets each form an equivalent channel matrix with a rank of $R = 10$, resulting in a QO-STBC for four transmit antennas with code length $T = 4$, code rate $= R/2T = 5/4$ and maximal symbolwise diversity. One of such solution sets is given in Appendix C and we will use it to discuss the relationship between equivalent channel matrix rank and code rate.

From the matrices $\mathbf{A}_1$ to $\mathbf{A}_{16}$ shown in Appendix C, one can easily verify that the matrices $\mathbf{A}_1$ to $\mathbf{A}_8$ satisfy the QOC with the matrices $\mathbf{A}_9$ to $\mathbf{A}_{16}$. In other words, they may form a QO-STBC with two quasi-orthogonal groups. However, although $\mathbf{A}_1$ to $\mathbf{A}_{16}$ are 16 different matrices, the equivalent channel matrix formed by them has a rank of only 10, instead of 16. In other words, only 10 out of these 16 matrices ($\mathbf{A}_1$ to $\mathbf{A}_5$ and $\mathbf{A}_9$ to $\mathbf{A}_{13}$) are linearly independent and can be used as dispersion matrices to carry information symbols, while the other 6 matrices ($\mathbf{A}_6$ to $\mathbf{A}_8$ and $\mathbf{A}_{14}$ to $\mathbf{A}_{16}$) are linearly dependent on the earlier 10 matrices and hence cannot be used to carry any new information symbol. Since every dispersion matrix carry a real information symbol and the matrices in Appendix C have length 4, the resultant QO-STBC has a code rate of 10/2/4 = 5/4.

Table I summarizes the results of our code research using various code parameters. It shows that rate-5/4 QO-STBCs exist with symbolwise diversity level = 4 and group = 2, while rate-4 QO-STBCs exist with symbolwise diversity level = 2 and group = 2. One of the solution sets found for the rate-4 QO-STBC is given in Appendix D. Other interesting observations include:

- All QO-STBCs with code rate greater than one (shaded rows in Table I) have dispersion matrices separated into 2 groups and weights greater than 1.

- To achieve a higher code rate from 5/4 to 4, the rank of the dispersion matrices (hence the symbolwise diversity level) is reduced from 4 to 2, i.e. full transmit diversity can no longer be achieved.

- The rate-4 diversity-2 length-4 QO-STBC found turns out to be equivalent to the rate-4 diversity-2 length-2 non-orthogonal STBCs [13, 15], and both of them have the same decoding complexity. Nonetheless its existence, when compared with the rate-5/4 code found, illustrates the tradeoff between code rate, diversity of high-rate QO-STBC.

## 4. Conclusion

In this paper, the algebraic constraint of the dispersion matrices of a QO-STBC is derived and applied in a graph-based approach to provide a computer search for QO-STBC with code rate greater than one. Due to the high complexity of this code search problem, an efficient Modified Depth First Search (MDFS) algorithm, which is extended from the Depth First Search algorithm well known in graph theory, is proposed to facilitate the code search. For 4 transmit antennas, a few QO-STBCs with rates 5/4 and 4 are found. A trade-off between the code rate and the symbolwise transmit diversity level is observed in these high-rate QO-STBCs, as the rate-5/4 codes have symbolwise diversity of 4 while the rate-4 codes have symbolwise diversity of 2. All the high-rate QO-STBCs obtained are also found to have dispersion matrices separable into 2 quasi-orthogonal groups and the "weights" of these dispersion matrices are found to be two, i.e. there are two non-zero entries in every row of the dispersion matrices. To our knowledge, the rate-5/4 codes are the first known non-trivial QO-STBCs with code rate > 1.

## AKNOWLEDGEMENTS

The authors would like to thank Hwor Shen Chong and Chee Muah Lim from Nanyang Technological University (Singapore) in assisting in part of this work. The authors would also like to thank the editor Dr Olav Tirkkonen and the anonymous reviewer for many constructive suggestions that help improve the quality of this paper.

**Table**

Table I Code search results found using proposed MDFS algorithm

| Parameters | | | |
|---|---|---|---|
| Matrix Rank (symbolwise diversity) | Matrix Weight | Group | Max. Code Rate |
| 4 | 1 | 4 | 1 |
| 4 | 1 | 2 | 1 |
| 4 | 2 | 2 | 5/4 |
| 2 | 2 | 8 | 1 |
| 2 | 2 | 2 | 4 |

**Figures**

$$\begin{bmatrix} 1 & 1 \\ 1 & -1 \end{bmatrix} \begin{bmatrix} 1 & 1 \\ -1 & 1 \end{bmatrix} \begin{bmatrix} 1 & -1 \\ 1 & 1 \end{bmatrix} \begin{bmatrix} -1 & 1 \\ 1 & 1 \end{bmatrix}$$

$$\begin{bmatrix} 1 & 1 \\ j & -j \end{bmatrix} \begin{bmatrix} 1 & 1 \\ -j & j \end{bmatrix} \begin{bmatrix} 1 & -1 \\ j & j \end{bmatrix} \begin{bmatrix} -1 & 1 \\ j & j \end{bmatrix}$$

$$\begin{bmatrix} 1 & j \\ 1 & -j \end{bmatrix} \begin{bmatrix} 1 & j \\ -1 & j \end{bmatrix} \begin{bmatrix} 1 & -j \\ 1 & j \end{bmatrix} \begin{bmatrix} -1 & j \\ 1 & j \end{bmatrix}$$

$$\begin{bmatrix} 1 & j \\ j & 1 \end{bmatrix} \begin{bmatrix} 1 & -j \\ j & -1 \end{bmatrix} \begin{bmatrix} 1 & -j \\ -j & 1 \end{bmatrix} \begin{bmatrix} 1 & j \\ -j & -1 \end{bmatrix}$$

Figure 1 Complex Hadamard matrices of weight two

$$\begin{bmatrix} \mathbf{P} & 0 \\ 0 & \mathbf{Q} \end{bmatrix} \begin{bmatrix} \mathbf{P} & 0 \\ 0 & -\mathbf{Q} \end{bmatrix} \begin{bmatrix} \mathbf{P} & 0 \\ 0 & j\mathbf{Q} \end{bmatrix} \begin{bmatrix} \mathbf{P} & 0 \\ 0 & -j\mathbf{Q} \end{bmatrix}$$

$$\begin{bmatrix} j\mathbf{P} & 0 \\ 0 & j\mathbf{Q} \end{bmatrix} \begin{bmatrix} j\mathbf{P} & 0 \\ 0 & -j\mathbf{Q} \end{bmatrix} \begin{bmatrix} j\mathbf{P} & 0 \\ 0 & -\mathbf{Q} \end{bmatrix} \begin{bmatrix} j\mathbf{P} & 0 \\ 0 & \mathbf{Q} \end{bmatrix}$$

$$\begin{bmatrix} 0 & \mathbf{P} \\ \mathbf{Q} & 0 \end{bmatrix} \begin{bmatrix} 0 & \mathbf{P} \\ -\mathbf{Q} & 0 \end{bmatrix} \begin{bmatrix} 0 & \mathbf{P} \\ j\mathbf{Q} & 0 \end{bmatrix} \begin{bmatrix} 0 & \mathbf{P} \\ -j\mathbf{Q} & 0 \end{bmatrix}$$

$$\begin{bmatrix} 0 & j\mathbf{P} \\ j\mathbf{Q} & 0 \end{bmatrix} \begin{bmatrix} 0 & j\mathbf{P} \\ -j\mathbf{Q} & 0 \end{bmatrix} \begin{bmatrix} 0 & j\mathbf{P} \\ -\mathbf{Q} & 0 \end{bmatrix} \begin{bmatrix} 0 & j\mathbf{P} \\ \mathbf{Q} & 0 \end{bmatrix}$$

Figure 2 Patterns to generate matrices of rank four and weight two

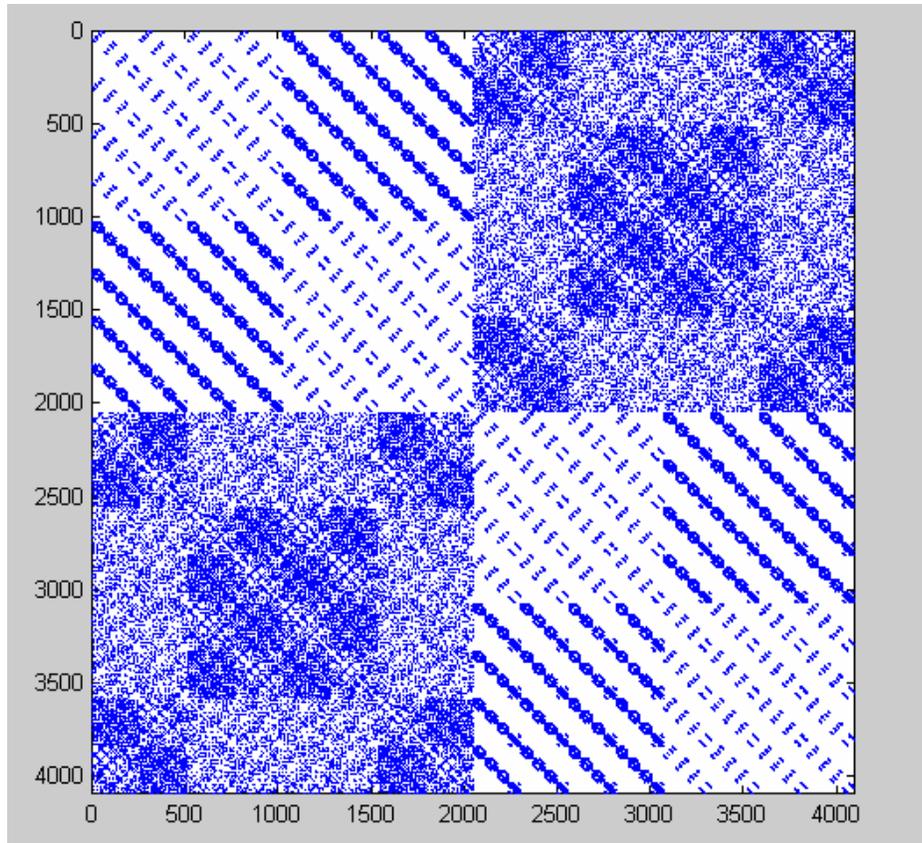

Figure 3 QOC link connection between 4×4 matrices formed in Step (a)

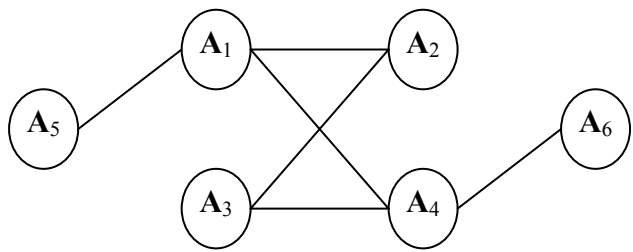

Figure 4 Example of a graph

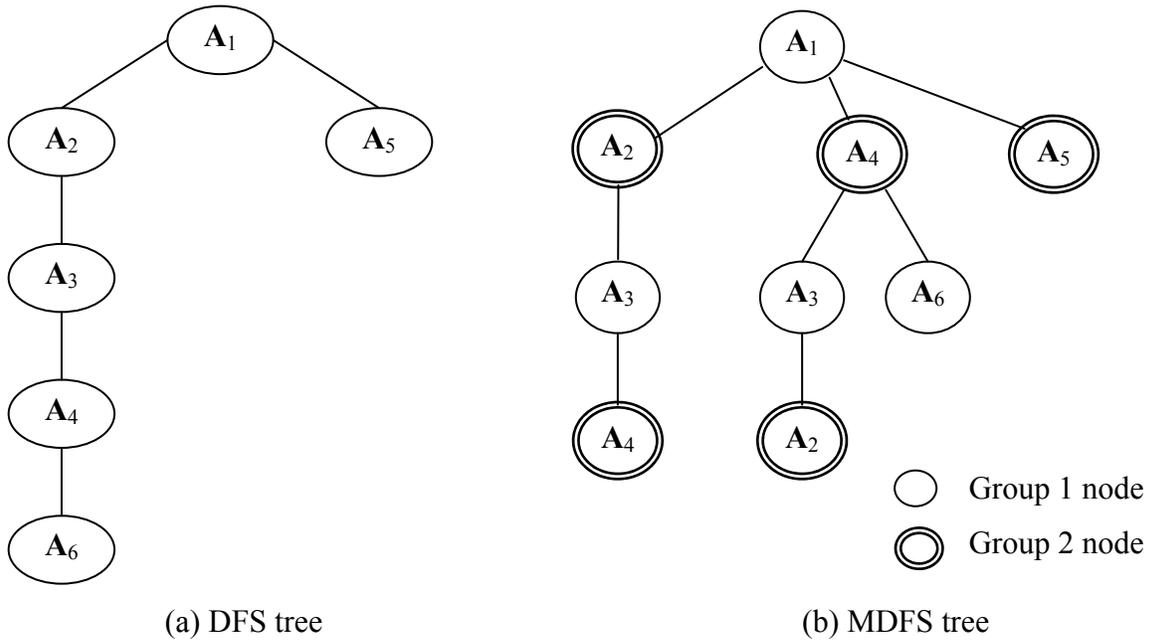

(a) DFS tree  (b) MDFS tree

Figure 5 Trees generated by DFS and MDFS algorithms

# APPENDIX A: Depth First Search (DFS) Algorithm

Input: Graph, number of nodes in graph ($N$)

Output: a spanning tree $T$ with every node visited only once

**Begin**

    Pick a node $n$, $n \leq N$ as starting point;

    Assign node $n$ as the root of tree $T$, i.e. $T = T \cup \{n\}$;

    DFS($n$, $T$);

**End**

**Procedure DFS($n$, $T$)**

**Begin**

    for ($v \in \{$neighbor of $n\}$)

        if ($v \notin T$)                   % i.e. $v$ not in the tree $T$

            Add node $v$ to the tree $T$ with $n$ as parent, i.e. $T = T \cup \{v\}$;

            DFS($v$, $T$);

        end

    end

**End**

## APPENDIX B: Modified Depth First Search (MDFS) Algorithm

Input: Graph, number of nodes in graph ($N$), number of groups required ($G$)

Output: a tree $T$ with every branch as a valid solution

**Begin**

    Pick a node $n$, $n \leq N$ as starting point;

    Assign node $n$ to Group 1, i.e. $g(n) = 1$;

    Assign node $n$ as the root of tree $T$, i.e. $T = T \cup \{n\}$;

    MDFS($n$, $T$, $g$);

**End**

**Procedure MDFS($n$, $T$, $g$)**

**Begin**

    for ($v \in \{$neighbor of $n\}$)

        if ($v \notin \{$ancestor of $n\}$)

            Assume that node $v$ is in Group $p$

            where $p = g(n)+1$ if $g(n)+1 \leq G$,

                    $= 1$     if $g(n)+1 > G$;

            if ($v$ possesses a link with all ancestors of $n$ with

               different groups, i.e. g(ancestor of n) $\neq p$)

                  Assign node $v$ to Group $p$, i.e. $g(v) = p$;

                  Add node $v$ to the tree $T$ with $n$ as parent, i.e. $T = T \cup \{v\}$;

                  MDFS($v$, $T$, $g$);

            end

        end

    end

**END**

# APPENDIX C: Matrices found by MDFS using $G = 2$, $T = N_t = 4$, rank = 4, weight = 2

*Group 1*

$$\mathbf{A}_1 = \begin{bmatrix} 1 & 1 & 0 & 0 \\ 1 & -1 & 0 & 0 \\ 0 & 0 & 1 & 1 \\ 0 & 0 & 1 & -1 \end{bmatrix}, \quad \mathbf{A}_2 = \begin{bmatrix} 1 & 1 & 0 & 0 \\ 1 & -1 & 0 & 0 \\ 0 & 0 & -1 & -1 \\ 0 & 0 & -1 & 1 \end{bmatrix}, \quad \mathbf{A}_3 = \begin{bmatrix} 1 & 1 & 0 & 0 \\ 1 & -1 & 0 & 0 \\ 0 & 0 & j & j \\ 0 & 0 & -j & j \end{bmatrix},$$

$$\mathbf{A}_4 = \begin{bmatrix} 1 & 1 & 0 & 0 \\ 1 & -1 & 0 & 0 \\ 0 & 0 & j & -j \\ 0 & 0 & j & j \end{bmatrix}, \quad \mathbf{A}_5 = \begin{bmatrix} 1 & 1 & 0 & 0 \\ 1 & -1 & 0 & 0 \\ 0 & 0 & -j & j \\ 0 & 0 & j & j \end{bmatrix}, \quad \mathbf{A}_6 = \begin{bmatrix} 1 & 1 & 0 & 0 \\ 1 & -1 & 0 & 0 \\ 0 & 0 & -j & -j \\ 0 & 0 & j & -j \end{bmatrix},$$

$$\mathbf{A}_7 = \begin{bmatrix} 1 & 1 & 0 & 0 \\ 1 & -1 & 0 & 0 \\ 0 & 0 & -j & j \\ 0 & 0 & -j & -j \end{bmatrix}, \quad \mathbf{A}_8 = \begin{bmatrix} 1 & 1 & 0 & 0 \\ 1 & -1 & 0 & 0 \\ 0 & 0 & j & -j \\ 0 & 0 & -j & -j \end{bmatrix}.$$

*Group 2*

$$\mathbf{A}_9 = \begin{bmatrix} -1 & 1 & 0 & 0 \\ 1 & 1 & 0 & 0 \\ 0 & 0 & -1 & 1 \\ 0 & 0 & 1 & 1 \end{bmatrix}, \quad \mathbf{A}_{10} = \begin{bmatrix} -1 & 1 & 0 & 0 \\ 1 & 1 & 0 & 0 \\ 0 & 0 & 1 & -1 \\ 0 & 0 & -1 & -1 \end{bmatrix}, \quad \mathbf{A}_{11} = \begin{bmatrix} j & j & 0 & 0 \\ j & -j & 0 & 0 \\ 0 & 0 & -1 & 1 \\ 0 & 0 & 1 & 1 \end{bmatrix},$$

$$\mathbf{A}_{12} = \begin{bmatrix} j & j & 0 & 0 \\ -j & j & 0 & 0 \\ 0 & 0 & -1 & 1 \\ 0 & 0 & 1 & 1 \end{bmatrix}, \quad \mathbf{A}_{13} = \begin{bmatrix} j & -j & 0 & 0 \\ j & j & 0 & 0 \\ 0 & 0 & -1 & 1 \\ 0 & 0 & 1 & 1 \end{bmatrix}, \quad \mathbf{A}_{14} = \begin{bmatrix} j & j & 0 & 0 \\ j & -j & 0 & 0 \\ 0 & 0 & 1 & -1 \\ 0 & 0 & -1 & -1 \end{bmatrix},$$

$$\mathbf{A}_{15} = \begin{bmatrix} j & j & 0 & 0 \\ -j & j & 0 & 0 \\ 0 & 0 & 1 & -1 \\ 0 & 0 & -1 & -1 \end{bmatrix}, \quad \mathbf{A}_{16} = \begin{bmatrix} j & -j & 0 & 0 \\ j & j & 0 & 0 \\ 0 & 0 & 1 & -1 \\ 0 & 0 & -1 & -1 \end{bmatrix}.$$

# APPENDIX D: Matrices found by MDFS using $G = 2$, $T = N_t = 4$, rank = 2, weight = 2

*Group 1*

$$\mathbf{A}_1 = \begin{bmatrix} 1 & 1 & 0 & 0 \\ 1 & 1 & 0 & 0 \\ 0 & 0 & 1 & 1 \\ 0 & 0 & 1 & 1 \end{bmatrix} \quad \mathbf{A}_2 = \begin{bmatrix} j & j & 0 & 0 \\ j & j & 0 & 0 \\ 0 & 0 & j & j \\ 0 & 0 & j & j \end{bmatrix} \quad \mathbf{A}_3 = \begin{bmatrix} 0 & 0 & 1 & 1 \\ 0 & 0 & 1 & 1 \\ 1 & 1 & 0 & 0 \\ 1 & 1 & 0 & 0 \end{bmatrix} \quad \mathbf{A}_4 = \begin{bmatrix} 0 & 0 & j & j \\ 0 & 0 & j & j \\ j & j & 0 & 0 \\ j & j & 0 & 0 \end{bmatrix}$$

$$\mathbf{A}_5 = \begin{bmatrix} 1 & 1 & 0 & 0 \\ 1 & 1 & 0 & 0 \\ 0 & 0 & -1 & -1 \\ 0 & 0 & -1 & -1 \end{bmatrix} \quad \mathbf{A}_6 = \begin{bmatrix} j & j & 0 & 0 \\ j & j & 0 & 0 \\ 0 & 0 & -j & -j \\ 0 & 0 & -j & -j \end{bmatrix} \quad \mathbf{A}_7 = \begin{bmatrix} 0 & 0 & 1 & 1 \\ 0 & 0 & 1 & 1 \\ -1 & -1 & 0 & 0 \\ -1 & -1 & 0 & 0 \end{bmatrix} \quad \mathbf{A}_8 = \begin{bmatrix} 0 & 0 & j & j \\ 0 & 0 & j & j \\ -j & -j & 0 & 0 \\ -j & -j & 0 & 0 \end{bmatrix}$$

$$\mathbf{A}_9 = \begin{bmatrix} -1 & 1 & 0 & 0 \\ -1 & 1 & 0 & 0 \\ 0 & 0 & -1 & 1 \\ 0 & 0 & -1 & 1 \end{bmatrix} \quad \mathbf{A}_{10} = \begin{bmatrix} -j & j & 0 & 0 \\ -j & j & 0 & 0 \\ 0 & 0 & -j & j \\ 0 & 0 & -j & j \end{bmatrix} \quad \mathbf{A}_{11} = \begin{bmatrix} 0 & 0 & -1 & 1 \\ 0 & 0 & -1 & 1 \\ 1 & -1 & 0 & 0 \\ 1 & -1 & 0 & 0 \end{bmatrix} \quad \mathbf{A}_{12} = \begin{bmatrix} 0 & 0 & j & -j \\ 0 & 0 & j & -j \\ -j & j & 0 & 0 \\ -j & j & 0 & 0 \end{bmatrix}$$

$$\mathbf{A}_{13} = \begin{bmatrix} 1 & -1 & 0 & 0 \\ 1 & -1 & 0 & 0 \\ 0 & 0 & -1 & 1 \\ 0 & 0 & -1 & 1 \end{bmatrix} \quad \mathbf{A}_{14} = \begin{bmatrix} j & -j & 0 & 0 \\ j & -j & 0 & 0 \\ 0 & 0 & -j & j \\ 0 & 0 & -j & j \end{bmatrix} \quad \mathbf{A}_{15} = \begin{bmatrix} 0 & 0 & 1 & -1 \\ 0 & 0 & 1 & -1 \\ 1 & -1 & 0 & 0 \\ 1 & -1 & 0 & 0 \end{bmatrix} \quad \mathbf{A}_{16} = \begin{bmatrix} 0 & 0 & j & -j \\ 0 & 0 & j & -j \\ j & -j & 0 & 0 \\ j & -j & 0 & 0 \end{bmatrix}$$

*Group 2*

$$\mathbf{A}_{17} = \begin{bmatrix} -1 & 1 & 0 & 0 \\ 1 & -1 & 0 & 0 \\ 0 & 0 & -1 & 1 \\ 0 & 0 & 1 & -1 \end{bmatrix} \quad \mathbf{A}_{18} = \begin{bmatrix} -j & j & 0 & 0 \\ j & -j & 0 & 0 \\ 0 & 0 & -j & j \\ 0 & 0 & j & -j \end{bmatrix} \quad \mathbf{A}_{19} = \begin{bmatrix} 0 & 0 & 1 & -1 \\ 0 & 0 & -1 & 1 \\ -1 & 1 & 0 & 0 \\ 1 & -1 & 0 & 0 \end{bmatrix} \quad \mathbf{A}_{20} = \begin{bmatrix} 0 & 0 & j & -j \\ 0 & 0 & -j & j \\ -j & j & 0 & 0 \\ j & -j & 0 & 0 \end{bmatrix}$$

$$\mathbf{A}_{21} = \begin{bmatrix} -1 & -1 & 0 & 0 \\ 1 & 1 & 0 & 0 \\ 0 & 0 & -1 & -1 \\ 0 & 0 & 1 & 1 \end{bmatrix} \quad \mathbf{A}_{22} = \begin{bmatrix} -j & -j & 0 & 0 \\ j & j & 0 & 0 \\ 0 & 0 & -j & -j \\ 0 & 0 & j & j \end{bmatrix} \quad \mathbf{A}_{23} = \begin{bmatrix} 0 & 0 & -1 & -1 \\ 0 & 0 & 1 & 1 \\ 1 & 1 & 0 & 0 \\ -1 & -1 & 0 & 0 \end{bmatrix} \quad \mathbf{A}_{24} = \begin{bmatrix} 0 & 0 & -j & -j \\ 0 & 0 & j & j \\ j & j & 0 & 0 \\ -j & -j & 0 & 0 \end{bmatrix}$$

$$\mathbf{A}_{25} = \begin{bmatrix} 1 & -1 & 0 & 0 \\ -1 & 1 & 0 & 0 \\ 0 & 0 & -1 & 1 \\ 0 & 0 & 1 & -1 \end{bmatrix} \quad \mathbf{A}_{26} = \begin{bmatrix} j & -j & 0 & 0 \\ -j & j & 0 & 0 \\ 0 & 0 & -j & j \\ 0 & 0 & j & -j \end{bmatrix} \quad \mathbf{A}_{27} = \begin{bmatrix} 0 & 0 & 1 & -1 \\ 0 & 0 & -1 & 1 \\ 1 & -1 & 0 & 0 \\ -1 & 1 & 0 & 0 \end{bmatrix} \quad \mathbf{A}_{28} = \begin{bmatrix} 0 & 0 & j & -j \\ 0 & 0 & -j & j \\ j & -j & 0 & 0 \\ -j & j & 0 & 0 \end{bmatrix}$$

$$\mathbf{A}_{25} = \begin{bmatrix} 1 & 1 & 0 & 0 \\ -1 & -1 & 0 & 0 \\ 0 & 0 & -1 & -1 \\ 0 & 0 & 1 & 1 \end{bmatrix} \quad \mathbf{A}_{26} = \begin{bmatrix} j & j & 0 & 0 \\ -j & -j & 0 & 0 \\ 0 & 0 & -j & -j \\ 0 & 0 & j & j \end{bmatrix} \quad \mathbf{A}_{27} = \begin{bmatrix} 0 & 0 & 1 & 1 \\ 0 & 0 & -1 & -1 \\ 1 & 1 & 0 & 0 \\ -1 & -1 & 0 & 0 \end{bmatrix} \quad \mathbf{A}_{28} = \begin{bmatrix} 0 & 0 & j & j \\ 0 & 0 & -j & -j \\ j & j & 0 & 0 \\ -j & -j & 0 & 0 \end{bmatrix}$$